\documentclass[prd,showpacs,preprintnumbers,amsmath,amssymb,floatfix]{revtex4}

\usepackage{graphicx}
\usepackage{dcolumn}
\usepackage{bm}
\usepackage{amssymb}
\usepackage{epsfig}
\usepackage{color}

\newcommand{\ba}{\begin{eqnarray}}
\newcommand{\ea}{\end{eqnarray}}
\newcommand{\be}{\begin{equation}}
\newcommand{\ee}{\end{equation}}
\newcommand{\bdisplay}{\begin{displaymath}}
\newcommand{\edisplay}{\end{displaymath}}

\newcommand{\eq}[1]{Eq.\,(\ref{#1})}

\begin{document}

\title{ Connection of the virtual $\gamma^*p$ cross section of $ep$ deep inelastic scattering to real $\gamma p$ scattering, and the implications for $\nu N$ and $ep$ total cross sections}

\author{ Martin~M.~Block}
\email{mblock@northwestern.edu}
\affiliation{Department of Physics and Astronomy, Northwestern University,
Evanston, IL 60208}
\author{Loyal Durand}
\email{ldurand@hep.wisc.edu}
\altaffiliation{Mailing address: 415 Pearl Ct., Aspen, CO 81611}
\affiliation{Department of Physics, University of Wisconsin, Madison, WI 53706}
\author{Phuoc Ha}
\email{pdha@towson.edu}
\affiliation{Department of Physics, Astronomy and Geosciences, Towson University, Towson, MD 21252}

\begin{abstract}
We show that it is possible to fit  all of the HERA DIS (deep inelastic scattering) data on $F_2^{\gamma p}$ at  small values of Bjorken $x$, including the data at  {\em very low} $Q^2$, using a new model for $F_2^{\gamma p}$ which both includes an asymptotic (high energy) part that satisfies a saturated Froissart bound behavior,  with  a vector-dominance like mass factor in the parameterization, and extends smoothly to $Q^2=0$.  We require that the corresponding part of the virtual $\gamma^* p$ cross section match the known asymptotic part of the real $\gamma p$ cross section at $Q^2=0$, a cross section which  is determined by strong interactions and asymptotically satisfies a saturated Froissart bound of the form $\alpha +\beta\ln s+\gamma\ln^2s$. Using this model for the asymptotic part of $F_2^{\gamma p}$ plus  a known valence contribution, we fit the asymptotic high energy part of the HERA data  with $x\le 0.1$ and $W\ge 25$ GeV; the fit is excellent. We find that the mass parameter in the fit lies in the region of the light vector mesons, somewhat above  the $\rho$ meson mass, and is compatible with vector dominance.  We use this fit to obtain accurate results for the  high energy $ep$ and isoscalar $\nu N$ total cross sections.  Both cross sections obey an analytic expression of the type $a +b \ln E +c \ln^2 E +d \ln^3 E$ at large energies $E$ of the incident particle, reflecting the fact that the underlying strong interaction parts of the  $\gamma^*p$, $Z^*N$ and $W^*N$  cross sections  satisfy the saturated Froissart bound.  Since approximately 50\% of the $\nu N$ center of mass (cms) energy is found in $W$---the cms energy of the strongly interacting intermediate vector boson-nucleon system---a study of ultra-high-energy neutrino-nucleon cross sections would allow us, for the first time, to explore {\em strong interactions at incredibly high energies}.

\end{abstract}

\pacs{ 12.15.Hh, 13.15.+g, 13.60.Hb, 96.50.S}

\maketitle


\section{Introduction \label{sec:introduction}}

The experimental program at the $ep$ collider HERA at the DESY laboratory studied deep inelastic sacttering (DIS) at small values of the Bjorken scaling variable $x$, defined in terms of the proton momentum $p$ and the electron momentum transfer $q$ in the scattering by $x=Q^2/2p\cdot q$. The measurements of the ZEUS \cite{ZEUS1,ZEUS2,ZEUS3} and H1 \cite{H1} detector groups covered the range $10^{-6}\lesssim x\lesssim 0.1$, with values of $Q^2=-q^2$ in the range of 0.1 GeV$^2$ to 5000 GeV$^2$, and determined the DIS structure function $F_2^{\gamma p}$ accurately in this region. This process can be viewed quite usefully as the scattering of a virtual photon $\gamma^*$ emitted by the electron on the target proton with $Q^2$, the negative (mass)$^2$ of the virtual photon, a measure of the virtuality of the process.  

It has been argued in a series of papers \cite{bbt1,bbt2,bhm,bdhmapp,bdhmFroissart} that the DIS structure function $F_2^{\gamma p}$ is basically hadronic in nature, reflecting the strong hadronic interactions initiated by the $\gamma^*$, and should show the saturated Froissart bounded  behavior \cite{froissart,martin1,martin2,martin3} observed for other hadronic scattering processes \cite{blockrev,mbair,blockhalzen} and real $\gamma p$ scattering \cite{BHgamma-p,blockaspen}.  The arguments are summarized in Ref.  \cite{bdhmFroissart}.  In particular, we note that as $Q^2\rightarrow 0$, the $\gamma^*p$ cross section should connect smoothly to the real $\gamma p$ cross section, for which Froissart-bounded behavior has been observed \cite{BHgamma-p}. 

In this communication, we will break up the structure function $F_2^{\gamma p}$ into two parts, a part that corresponds to high hadronic energies $W\equiv {\sqrt s}= \sqrt{(p+q)^2}$ in the final state and asymptotically satisfies the saturated Froissart bound, and a low energy part due to valence quarks. We will require that the asymptotic cross section  for $\gamma^* p$ scattering, $\sigma^{\gamma^* p}(W, Q^2)$ go smoothly  to the asymptotic cross section for real $\gamma p$ scattering, $\sigma^{\gamma p}(W)$, as $Q^2\rightarrow 0$. We will then add back in the low energy contribution of the valence quarks and do a 9 parameter $\chi^2$ fit to 395 HERA $F_2^{\gamma p}$ datum points with $W\ge 25$ GeV , $x\le 0.1$, and $0.15\le Q^2\le 3000$ GeV$^2$, an enormous range.

We then apply these results to determine the total high energy cross sections for the scattering processes $e+p\rightarrow e+ X$ and $\nu +N\rightarrow \ell +X$ where $X$ is a hadronic system containing any number of particles, including at least one nucleon, and $N$ is the isoscalar nucleon target $N=(n+p)/2$. Both cross sections have the same analytic form $\sigma = a +b \ln E + c \ln^2 E+d\ln^3 E$  when $E$, the laboratory energy of the incident electron or neutrino, is sufficiently high.  

We show further in the case of neutrinos that a large fraction of the center-of-mass (cms) energy of the $\nu N$ system appears in $W$, the cms energy of the final {\em hadronic system}.  Thus, by measuring this cross section at ultra-high (cosmic ray) neutrino energies, we would be able to explore hitherto unavailable hadronic energies.


\section{Continuation  of the virtual $\gamma^*\,p$ cross section to $Q^2=0$ \label{sec:continuation}}

\subsection{Definition of $\gamma^*\,p$ cross section\label{subsec:definition_of_sigma}}


The total inelastic cross section for the scattering of  a virtual photon $\gamma^*$ with 4-momentum $q$ on the proton with 4-momentum $p$ in deep inelastic $ep$ scattering (DIS), $\gamma^*\,p\rightarrow X$, is given by
\label{sigma_defined}
\ba
\label{sigma_F2}
\sigma^{\gamma * p}(s,Q^2) &=& \frac{16\pi^2\alpha}{\cal F}\left(1+\frac{\nu^2}{Q^2}\right) mW_2^{\gamma p}(s,Q^2) \nonumber \\
&=& \frac{8\pi^2\alpha}{\cal F}\left(1+\frac{2mx}{\nu}\right) \frac{1}{x}F_2^{\gamma p}(s,Q^2).
\ea
Here $F_2^{\gamma p}=\nu W_2^{\gamma p}$ is the usual DIS structure function,  $Q^2=-q^2$, $2m\nu=2p\cdot q$ with $\nu$ the energy of the photon in the proton rest frame, $x$ is the Bjorken scaling variable,  $x=Q^2/2p\cdot q\leq 1$,   $s=W^2=(p+q)^2=2p\cdot q-Q^2+m^2$ where $W$ is the total energy of the final hadronic system $X$, and $\cal F$ is the  flux factor. At high energies, $\nu\gg m$ and  the term $2mx/\nu$ in \eq{sigma_F2} can be dropped.

This definition is arbitrary to the extent that the  $\gamma^*p$ flux factor $\cal F$ cannot be derived in the usual way for virtual photons with $|{\bf Q}|>Q_0$, so some choice must be made subject to the condition that $\cal F$ reduce to the result for a real photon  for $Q^2\rightarrow 0$, i.e., ${\cal F}\rightarrow 4m\nu= 2(s-m^2)$, where $m$ is the mass of the proton. The continuation of the usual flux factor ${\cal F}=4\sqrt{(p_1\cdot p_2)^2-p_1^2p_2^2}=4\sqrt{(p\cdot q)^2-m^2q^2}$ to $q^2=-Q^2<0$ has this property, and gives 
\ba
{\cal F} &=& 4\left[(p\cdot q)^2+m^2Q^2\right]^{1/2} \nonumber \\
&\approx& 4p\cdot q=2(s+Q^2-m^2 ) 
\label{flux_factor}
\ea
at high energies, $\nu^2\gg Q^2$ or $2mx/\nu\ll 1$; we will use this definition. The so-called Hand convention used in some calculations corresponds to the alternative choice
\be
\label{Hand_K}
{\cal F}=2(s-m^2).
\ee
The difference at high energies is just in the neglect of the $Q^2$ in \eq{flux_factor}. This only makes a difference for $Q^2$ similar in size to $s$, that is, for the Bjorken variable $x$ near 1; the difference is unimportant in the large $W$, small $x$ region with which we will mainly be concerned. The arbitrariness in $\cal F$ does not, in any case, affect the fit to $F_2^{\gamma p}$ discussed below.\footnote{The only place the arbitrariness in $\cal F$ affects our results is in Fig.\ \ref{figure:gamP-sigma}, where we plot $\sigma^{\gamma^*p}$ as a function of $W$ at fixed values of $Q^2$, and then, given the cutoff $s=W^2\geq 625$ GeV$^2$ used in the fit, only at the higher values of $Q^2$ where the curves and datum points would be shifted somewhat if we were to use the Hand convention.}

Using the definition of ${\cal F}$ in \eq{flux_factor} the result for the $\gamma^*$-$p$ cross section at high energies is
\be
\label{sigma_F2_relation}
\sigma^{\gamma * p}(s,Q^2) = \frac{4\pi^2\alpha}{Q^2}F_2^{\gamma p}(s,Q^2),
\ee
with only a small change for the Hand convention. The structure function $F_2^{\gamma p}$ and the contribution of longitudinally polarized virtual photons to the cross section vanish in the limit $Q^2\rightarrow 0$, so the cross section is non-singular and involves only real, transversely polarized photons in this limit. 

The problem, then, in connecting the virtual $\gamma^*\,p$ cross sections determined from DIS experiments to the real $\gamma\,p$ cross sections measured at HERA and lower energies, is to find a form of $F_2^{\gamma p}$ that is consistent with the DIS data, and that has the right properties to  extrapolate smoothly to  $Q^2=0$ to match the measured $\gamma\,p$ data.


\subsection{Extrapolation of $F_2^{\gamma p}$ and the $\gamma^*\,p$ cross section to $Q^2=0$ \label{subsec:extrapolation}}

In \cite{bhm,bdhmapp} we presented an accurate fit to the HERA data for $Q^2\geq 0.85$ GeV$^2$, $x\lesssim 0.1$, using the data as combined by the ZEUS and H1 groups \cite{HERAcombined} and the  Froissart-bounded model proposed by Berger, Block, and Tan \cite{bbt2},
\ba
 F_2^{\gamma p}(x,Q^2) &=& (1-x)\left[\frac{F_P}{1-x_P}+A_1(Q^2)\ln\left(\frac{x_P}{x}\frac{1-x}{1-x_P}\right)  \right. \nonumber \\
                    &  & \left. +A_2(Q^2)\ln^2\left(\frac{x_P}{x}\frac{1-x}{1-x_P}\right)\right].
\label{bbtmodel}
 \ea
In this expression $A_1$ and $A_2$ are quadratic polynomials in $\ln{Q^2}$ and $x_P$, $F_P$ are the location of the approximate fixed point in $F_2^{\gamma p}$ observed in the data, and the value of $F_2^{\gamma p}$ at that point. The factor $(1-x)/x$ in the argument of the logarithms is just $s/Q^2$, where $s=(p+q)^2$ is the total energy in the $\gamma^*p$ rest frame; this is normalized to its value $(1-x_P)/x_P$ at the fixed point. This form for $F_2^{\gamma p}$ behaves asymptotically as $\ln^2s$ as $x$ decreases at fixed $Q^2$, reflecting the expected Froissart-bounded behavior, and describes the HERA data well. 

Unfortunately, the model does not have the properties necessary for the $\gamma^*p$ cross section defined in  \eq{sigma_F2_relation} to extend smoothly to $Q^2=0$ at fixed $s$ to connect with the real $\gamma p$ cross section:  $F_2^{\gamma p}$ does not vanish for $Q^2\rightarrow 0$, but rather diverges  as powers of $\ln{Q^2}$ through the coefficient functions $A_1$ and $A_2$ and the terms in $\ln(1/x)=\ln[(s-Q^2+m^2)/Q^2]$. It also behaves badly phenomenologically in the valence region and with respect to the HERA data for $Q^2\lesssim 1$ GeV$^2$.

To motivate the modified expression we will use to extend $F_2^{\gamma p}$ and $\sigma^{\gamma^* p}(W,Q^2)$ to $Q^2=0$ for large $W$, we recall that Ashok suri \cite{ashok} showed that the $\gamma^*$-$p$ cross section, expressed as the imaginary part of the forward $\gamma^*$-$p$ scattering amplitude, is real analytic and satisfies standard single-variable dispersion relations simultaneously in the variables $Q^2$ and $\nu = p\cdot q/m$ (but not in $Q^2$ with the choice of variables $Q^2$,  $s$). The dispersion relation in $Q^2$ holds for $Q^2<4m_\pi^2$ for physical values of $\nu$; $F_2^{\gamma p}\propto Q^2\,\sigma^{\gamma p}$ satisfies a similar dispersion relation, and vanishes for $Q^2\rightarrow 0$ at fixed $\nu$. As a result, we can write a once-subtracted dispersion relation in $Q^2$ for $F_2^{\gamma p}(\nu,Q^2)$,
\be
\label{F2_dispersn_reln}
F_2^{\gamma p}(\nu,Q^2) = Q^2\int_{4m_\pi^2}^\infty dz \frac{f_2(\nu,z)}{z(z+Q^2)},
\ee
where the weight function $f_2(\nu,z)$ is the absorptive part of $F_2^{\gamma p}$. For reasonably rapid convergence, the integral falls for large $Q^2$ as $\sim 1/Q^2$ up to a multiplicative factor that does not upset the convergence. The idea of vector meson dominance suggests that the absorptive part should be largest in the region of the lower-mass vector mesons. We will assume this is the case.

$F_2^{\gamma p}$ is still subject to the Froissart bound for $s=2m\nu-Q^2+m^2$ or $\nu$ tending toward $\infty$ at fixed $Q^2$ as assumed in \eq{bbtmodel} and \cite{bhm,bdhmapp}. Those models provide good fits to the HERA data at small $x$ and $Q^2$ above 1-2 GeV$^2$, so we would like our extended model to be compatible with them at large $Q^2$. However, it is important to distinguish contributions to $F_2^{\gamma p}$ which involve large hadronic energies $W$  from lower energy contributions; in the HERA region the latter appear through valence contributions to $F_2^{\gamma p}$, which we will therefore single out \footnote{The high- and low-$W$ regions were not distinguished in making the fits in \cite{bbt2,bhm,bdhmapp}; however, valence effects were treated approximately in \cite{bdhmapp} by joining a contribution with the valence shape smoothly to the fitted $F_2^{\gamma p}$ specified by \eq{bbtmodel}, truncated at a small value of $x$, and then checking that the quark and momentum sum rules were satisfied by the composite expression.}.

Given this input, we write $F_2^{\gamma p}$ as
\be
\label{F2_split}
 F_2^{\gamma p}(x,Q^2) = F_{2,{\rm v}}^{\gamma p}(x,Q^2)+F_{2,{\rm asymp}}^{\gamma p}(x,Q^2) ,
 \ee
 where $F_{2,{\rm v}}^{\gamma p}$ is a valence term and $F_{2,{\rm asymp}}^{\gamma p}$ is an asymptotic term which we assume has the Froissart-bounded form
\ba 
\label{F2asymp}
F_{2,{\rm asymp}}^{\gamma p}(x,Q^2) &=& D(Q^2)(1-x)^n \left[ C(Q^2)+ A(Q^2)\ln \left(\frac{1}{x}\frac{Q^2}{Q^2+\mu^2}\right)+B(Q^2) \ln^2 \left(\frac{1}{x}\frac{Q^2}{Q^2+\mu^2}\right) \right] \\
\label{F2asymp_nu}
&=& D(Q^2)(1-Q^2/2m\nu)^n \left[ C(Q^2)+ A(Q^2)\ln \left(\frac{2m\nu}{Q^2+\mu^2}\right)+B(Q^2) \ln^2 \left(\frac{2m\nu}{Q^2+\mu^2}\right) \right].
\ea

We take the coefficient function $A(Q^2),\,B(Q^2),\,C(Q^2)$ to have the logarithmic form at large $Q^2$ assumed in \cite{bbt2,bhm,bdhmapp} since this behavior describes the large-$Q^2$ HERA data well, but have modified the arguments to eliminate the divergences that appeared in \cite{bbt2}  for $Q^2\rightarrow 0$, taking
\ba
    A(Q^2)&=& a_0+a_1\,\ln \left(1+\frac{Q^2}{\mu^2}\right) +a_2\,\ln^2 \left(1+\frac{Q^2}{\mu^2}\right) , \nonumber \\
    B(Q^2)&=& b_0+b_1\,\ln \left(1+\frac{Q^2}{\mu^2}\right) +b_2\,\ln^2 \left(1+\frac{Q^2}{\mu^2}\right), \\
    C(Q^2)&=& c_0+c_1\,\ln \left(1+\frac{Q^2}{\mu^2}\right).  \nonumber \label{ABC}
\ea
Here $\mu^2$ is a scale factor which determines the transition to an asymptotic $\ln(Q^2/\mu^2)$ behavior of the logarithms in accordance with \eq{bbtmodel}.

In accord with the analysis of analytic properties of $F_2^{\gamma p}$ by Ashok suri \cite{ashok}, the arguments of the $x$-dependent logarithms have also been modified, with the variable $s/Q^2=(1-x)/x$ used in \eq{bbtmodel} and \cite{bbt2,bhm,bdhmapp} replaced by $2m\nu$. Absorbing powers of $\ln(Q^2+\mu^2)$ from the coefficient functions, we then write the argument of the energy-dependent logarithms as $2m\nu/(Q^2+\mu^2)$.  This approaches $1/x$ for $Q^2\gg \mu^2$, reproducing the dominant $1/x$ behavior at small $x$ in \eq{bbtmodel}, but approaches $2m\nu/\mu^2$ as $Q^2\rightarrow 0$, so is well defined  at $Q^2=0$ in contrast to $1/x=2m\nu/Q^2$. 

The remaining overall factor $D(Q^2)$ represents the residual effects of the dispersion relation in $Q^2$. For a pure vector dominance model, with $f_2(\nu,z)$ a sum of delta functions in $z$ at the masses of the light vector mesons, it would take the form $D(Q^2)=Q^2\sum_iR_i(\nu)/(Q^2+M_i^2)$ with the residues $R_i(\nu)$ energy dependent and bounded by $\ln^2\nu$.   We assume that the additional logarithms in the coefficient functions arise from corrections to this simple picture, with $f_2(\nu,z)$ in \eq{F2_dispersn_reln} a more complicated function of $z$. 

In a simple two-meson model, we can write $D(Q^2)$ as 
\be
\label{Dinitial}
D(Q^2) = \frac{Q^2(Q^2+M_3^2)}{(Q^2+M_1^2)(Q^2+M_2^2)},
\ee
where we have normalized $D(Q^2)$ to 1 for $Q^2\rightarrow\infty$ by adjusting the numerical coefficients in \eq{ABC}. The $M$'\,s in this expression are to be interpreted as effective masses. We have investigated this form in our fits, discussed below, and found that the HERA data are not sufficient to distinguish two or more different masses  in the denominator, with $M_1\approx M_2$, so will take this as an equality and rewrite $D$ in a form that involves only a single mass and an additional parameter $\lambda$ as
\be
\label{Dfinal}
D(Q^2) = \frac{Q^2(Q^2+\lambda M^2)}{(Q^2+M^2)^2}.
\ee
We would expect the effective mass $M$ to lie in the range of the light vector-meson masses, and $\mu^2$, representing the effects of a broadened distribution, to potentially be somewhat larger.


\section{A Froissart-bounded fit to the $\gamma p$  and HERA $\gamma^*p$ data \label{sec:fits_to_data}}


\subsection{The Block-Halzen fit to the $\gamma p$ cross section \label{subset:BlockHalzen_fit}}

Block and Halzen \cite{BHgamma-p} have given a careful analysis of $\gamma p$ scattering assuming that $\sigma^{\gamma p}$ can be described in the region   $W\gtrsim 2$ GeV as the sum of a Regge-like term that falls with increasing center-of-mass energy $W$ as $1/W$, and a rising Froissart-bounded term, i.e.,
\be
\label{sigmaBH}
\sigma^{\gamma p}_{BH} =  \beta\left(\frac{m}{\nu}\right)^{1/2} +c_0^{BH}+c_1^{BH}\ln\frac{\nu}{m}+c_2^{BH}\ln^2\frac{\nu}{m}, 
\ee
where $\nu$ is the laboratory $\gamma$ energy, $Q^2=0$, and $W^2\equiv 2 m \nu$.
The parameters in the Block-Halzen (BH) fit were constrained by requiring that it match smoothly to the very precise fit in the resonance region given by Damashek and Gilman \cite{damashek}. In the following, we will use the parameters given in BH fit 2; this fits the $\gamma p$ cross section data for $W>2.01$ GeV, including the high energy measurements of the H1 \cite{H1gamma-p} and ZEUS \cite{ZEUSgamma-p} groups at HERA at $W= 200$ and 209 GeV, with a fit probability of 0.88: $c_0^{BH}=92.5\pm 6.8\ \mu{\rm b}$, $c_1^{BH}=-0.46\pm 2.88\  \mu{\rm b}$, $c_2^{BH}=0.803\pm 0.273\  \mu{\rm b}$, $\beta=78.4\pm 9.1\ \mu{\rm b}$.

With the form of $F_2^{\gamma p}$ given in \eq{F2asymp_nu}, the extension of the asymptotic, high $W$, part of the $\gamma^*p$ cross section  to $Q^2=0$ is smooth and gives as the asymptotic part of the real $\gamma p$ cross section
\be 
\label{sigma_gamma-p_asymp}
\sigma^{\gamma p}_{\rm asymp} = \lambda \frac{4\pi^2\alpha}{M^2} \left[c_0+a_0 \ln \left(\frac{\nu}{m} \frac{2m^2}{\mu^2}\right) +b_0 \ln^2 \left(\frac{\nu}{m} \frac{2m^2}{\mu^2}\right) \right]. 
\ee
By requiring that the expression in \eq{sigma_gamma-p_asymp} reproduce the asymptotic part of the Block-Halzen fit to the $\gamma  p$ cross section, ignoring the Regge-like term, we find that
\ba
c_2^{BH} &=& \lambda \frac{4\pi^2\alpha}{M^2} b_0 \, , \nonumber \\
 \label{constraints}
c_1^{BH} &=& \lambda \frac{4\pi^2\alpha}{M^2} \left (a_0+2 b_0 \ln \frac{2m^2}{\mu^2} \, ,\right) \\
c_0^{BH} &=& \lambda \frac{4\pi^2\alpha}{M^2} \left(c_0+ a_0 \ln \frac{2m^2}{\mu^2}+ b_0 \ln^2 \frac{2m^2}{\mu^2}\right) \, . \nonumber
\ea

The form  for $F_2^{\gamma p}$ given in Eqs.\ (\ref{F2asymp_nu}) and (\ref{Dfinal}) involves 12 parameters: $\lambda$, $n$, $M$, $\mu$, $c_0$, $c_1$, $a_0$, $a_1$, $a_2$, $b_0$, $b_1$, and $b_2$. Three parameters can be eliminated using Eq. (\ref{constraints}). We chose to eliminate the parameters $M$, $\mu$, and $c_0$ by expressing them in terms of the other parameters, so have a 9 parameter model for $F_{2,{\rm asymp}}^{\gamma p}$.
%


\subsection{ A new Froissart-bounded fit to the HERA data at high energies \label{subsec:HERAfit}}

We used the model of $F_2^{\gamma p}$ in \eq{F2_split} to fit the  combined HERA data \cite{HERAcombined}  in a way which did not introduce a low-energy bias in the fit to the asymptotic Froissart-bounded term $F_{2,{\rm asymp}}^{\gamma p}$. Results from the Block-Halzen fit to the $\gamma p$ cross sections, and fits to a number of strong interaction cross sections \cite{blockrev,blockaspen}, indicate that the influence of resonance and falling Regge terms in the cross sections becomes small for hadronic center-of-mass energies $W\gtrsim 25$ GeV, and that the rising Froissart-bounded asymptotic contributions successfully fit and predict the cross sections at much larger $W$, notably to 57 TeV in the case of $pp$ scattering \cite{blockhalzen}. To emphasize high energies, we therefore restricted the HERA data used in the fit to those with $W = \sqrt{ Q^2 (1-x)/x}\ge 25$ GeV. We further imposed the condition $x\le 0.1$ to ensure that the asymptotic term is at least comparable in magnitude to the valence term in \eq{F2_split} in the region used in the fit. 

In our fitting procedure, we subtracted the valence contribution $F_{2,{\rm v}}^{\gamma p}$ from the HERA data and fit the remainders using only the asymptotic expression in \eq{F2asymp}. We took the valence term in \eq{F2_split} from the CTEQ6L \cite{CTEQ6L,web} parton distributions, using the Mathematica program Interpolation to interpolate among the listed values for  $1.69$ GeV$^2 \leq Q^2 \leq 10^5$ GeV$^2$ and $10^{-6} \leq x \leq 0.8$.  With the conditions imposed above, we did not run into problems for $Q^2$ less than the minimum value 1.69 GeV$^2$ used in the CTEQ6L analysis: $W\ge 25$ GeV with $Q^2\le1.69$ GeV$^2$ requires $x\le 0.0027$, and the valence term is much smaller than the uncertainties in $F_2^{\gamma p}$ for smaller $Q^2$ and $x$, and can be taken as zero.

Using the parameterization for $F_{2,{\rm asymp}}^{\gamma p}$ in \eq{F2asymp} with the three parameters $M^2$, $\mu$, and $c_0$ eliminated, we fit the valence-corrected HERA data \cite{HERAcombined} at 41 different values of $Q^2$ with $x\leq 0.1$, covering a large range
of $Q^2$, $0.15 \le Q^2\le 3000$ GeV$^2$ (i.e., data for $Q^2=$
0.15, 0.2, 0.25, 0.35, 0.4, 0.5, 0.65, 0.85, 1.2, 1.5, 2.0, 2.7, 3.5, 4.5, 6.5, 8.5, 10, 12, 15, 18, 22,
27, 35, 45, 60, 70, 90, 120, 150, 200, 250, 300, 400, 500, 650,
800, 1000, 1200, 1500, 2000, and 3000 GeV$^2$). 
This data set has a total of 395 datum points. The use of the
sieve algorithm \cite{sieve} with a cutoff ($\Delta \chi^2_{i, {\rm max}}=6.0$) to sift the data and eliminate possible outlying datum points eliminated 6 points whose total contribution to the initial $\chi^2$ was 47.55, 13\% of the total. 

 The values of the 9 fit parameters, along with their statistical errors, are given in Table \ref{table:results}.  The remaining parameters, calculated from the matching conditions for the $\gamma p$ cross section in \eq{constraints}, are $M^2$=0.753$\pm$0.068 GeV$^2$, $\mu^2$=2.82$\pm$0.29 GeV$^2$, and $c_0$=0.255$\pm$0.016. Note that the calculated value of $M = 0.87\pm 0.04$ GeV is in the region spanned by the $\rho,\,\omega$, and $\phi$   meson masses as expected for real vector dominance at low $Q^2$.
 
 This fit gives a minimum $\chi^2_{\rm min}=324.35$ for 380 degrees of freedom. Renormalizing $\chi^2$ by the sieve factor ${\cal R}=1.109$ \cite{sieve} to correct for the truncation of the distribution at $\Delta \chi^2_{i, {\rm max}}=6.0$ gives a corrected $\chi^2=359.87$ for 380 degrees of freedom, or 0.95 per degree of freedom. The chance of finding a larger  $\chi^2$ in a normal $\chi^2$ distribution is 0.76 , so the fit is excellent. This is not changed significantly if the outlying points are included, with $\chi_{\rm min}^2 =371.9$ for 386 degrees of freedom, and a fit probability of 0.69.\footnote{For comparison, the fit obtained previously \cite{bhm,bdhmapp} using the Berger-Block-Tan parametrization in \eq{bbtmodel} with 7 free parameters ($x_P$ was fixed by hand) gave a corrected ${\cal R}\chi_{\rm min}^2=391$ for 356 datum points with $Q^2\geq 0.85$ after the elimination of 14 outlying points using the sieve algorithm \cite{sieve}. This corresponds to  fit probability of 0.033. For the restricted data set with $Q^2\geq 2.7$ GeV$^2$ used in the later analysis in that paper, there were 303 datum points and 7 free parameters, with 3 outliers excluded, so 293 degrees of freedom, with a corrected ${\cal R}\chi_{\rm min}^2=323.6$. The fit probability is 0.11, a good fit for this much data, though not as good as that obtained here including {\em all} the data down to $Q^2=0.15$. As expected from  the form of \eq{F2asymp}, the fits are very similar in the high-$Q^2$ region. 
}  We note that our results, derived using the HERA data for $W>25$ GeV and all $Q^2\geq 0.15$ GeV$^2$, are equivalent or better in quality to the the 10 parameter parton-level fit HERAPDF1 obtained by the HERA groups \cite{HERAcombined} using their data at all $x$ or $W$, but with the restriction $Q^2\geq 3.5$ GeV$^2$; this gave a $\chi^2$ per degree of freedom of $574/582 = 0.99$ and a fit probability of 0.59.

\begin{table}[ht]                   
%
\begin{center}
\def\arraystretch{1.2}            
     \caption{\label{fitted}\protect\small Results of our 9-parameter fit to the valence-corrected
HERA data for $F_{2,{\rm asymp}}^{\gamma p}(x,Q^2)$,  \eq{F2asymp}, for $0.15 \le
Q^2\le 3000$ GeV$^2$, subject to the restrictions $W>25$ GeV, $x<0.1$. The parameters fixed by the Block-Halzen fit to the real $\gamma p$ cross section \cite{BHgamma-p} are $M^2=0.753\pm0.068$ GeV$^2$, $\mu^2=2.82\pm0.29$ GeV$^2$, and $c_0=0.255\pm0.016$. \label{table:results}}
\begin{tabular}[b]{|l||c||}
    \multicolumn{1}{l}{Parameters}&\multicolumn{1}{c} {Values}\\
\hline
      $a_0$&$\phantom{-}8.205\times 10^{-4}\pm 4.62\times 10^{-4}$ \\
      $a_1$&$-5.148\times 10^{-2}\pm 8.19\times 10^{-3}$\\
      $a_2$&$-4.725\times 10^{-3}\pm 1.01\times 10^{-3}$\\
\hline
    $b_0$ &$\phantom{-}2.217\times 10^{-3}\pm 1.42\times 10^{-4}$\\
      $b_1$&$\phantom{-}1.244\times 10^{-2}\pm 8.56\times 10^{-4}$\\
      $b_2$&$\phantom{-}5.958\times 10^{-4}\pm 2.32\times 10^{-4}$ \\
\hline
$c_1$&$\phantom{-}1.475\times 10^{-1}\pm 3.025\times 10^{-2}$\\
\hline
$n$&$\phantom{-}11.49 \pm 0.99$\\
$\lambda$&$\phantom{-}2.430 \pm 0.153$\\
\hline
    \cline{1-2}
        \hline
    \hline
    $\chi^2_{\rm min}$&324.35\\
    ${\cal R}\times\chi^2_{\rm min}$&359.87\\
    d.o.f.&380\\
\hline
	${\cal R}\times\chi^2_{\rm min}$/d.o.f.&0.95\\
\hline
\end{tabular}
\end{center}
\end{table}
\def\arraystretch{1}  

\begin{figure} [ht]
\begin{center}
\mbox{\epsfig{file=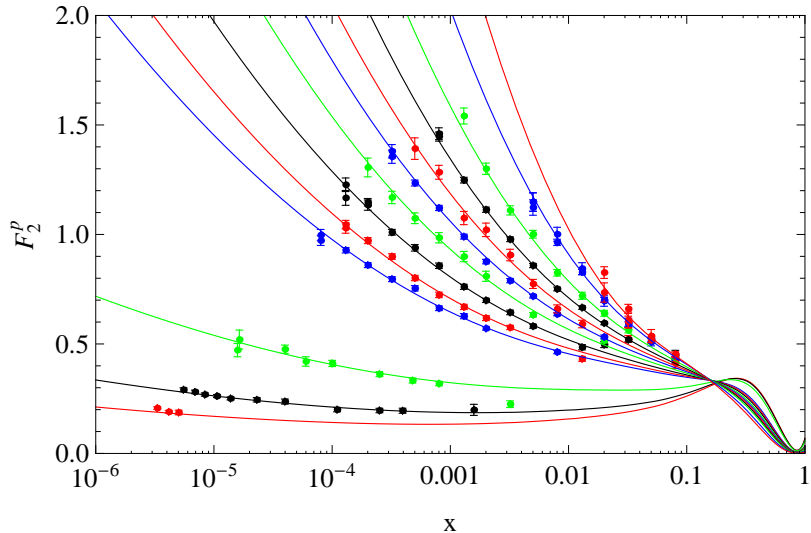,width=4.2in }}
\end{center}
\caption[]{Plots of the fitted proton structure function,
$F_2^{\gamma p}(x,Q^2)$ versus Bjorken $x$ for virtualities, bottom
to top, $Q^2$=0.15, 0.25, 0.65, 3.5, 4.5, 6.5, 10, 15, 22, 35, 70, 250, and
1200 GeV$^2$.} \label{figure:F2p}
\end{figure}

Adding the CTEQ6L valence contribution $F_{2v}$ to the $F_{2,{\rm asymp}}^{\gamma p}$ from the fit, we then obtain $F_2^{\gamma p}(x,Q^2)$.  Plots of the resulting $F_2^{\gamma p}(x,Q^2)$ versus $x$ are shown in Fig. \ref{figure:F2p} along with the all the corresponding (unsifted) HERA data for  representative values of $Q^2$, $Q^2$=0.15, 0.25, 0.65, 3.5, 4.5, 6.5, 10, 15, 22, 35, 70, 250, and 1200 GeV$^2$. 

Although not shown in the figure, we find that this fit and our published fit  \cite{bhm,bdhmapp} using the expression  in \eq{bbtmodel}---not valid for small $Q^2$---agree in the region where there are HERA data.  The present results  behave properly at smaller $Q^2$ and in the valence-dominated region $x\gtrsim 0.1$, which the earlier results did not. We also note the appearance of at quasi fixed point in the fit at $x\approx 0.11$ where the rise in the valence distribution with increasing $x$ approximately compensates for the sharp fall in the asymptotic distribution. This appeared automatically; it was imposed in \eq{bbtmodel}. 

The validity of this fit to $F_2^{\gamma p}$ could be checked in the future at the proposed Large Hadron-electron Collider (LHeC) \cite{LHeC} over ranges of $x$ and $Q^2$ larger by roughly a factor of 20 than those accessible at HERA.

%

\begin{figure} [htb]
\begin{center}
\mbox{\epsfig{file=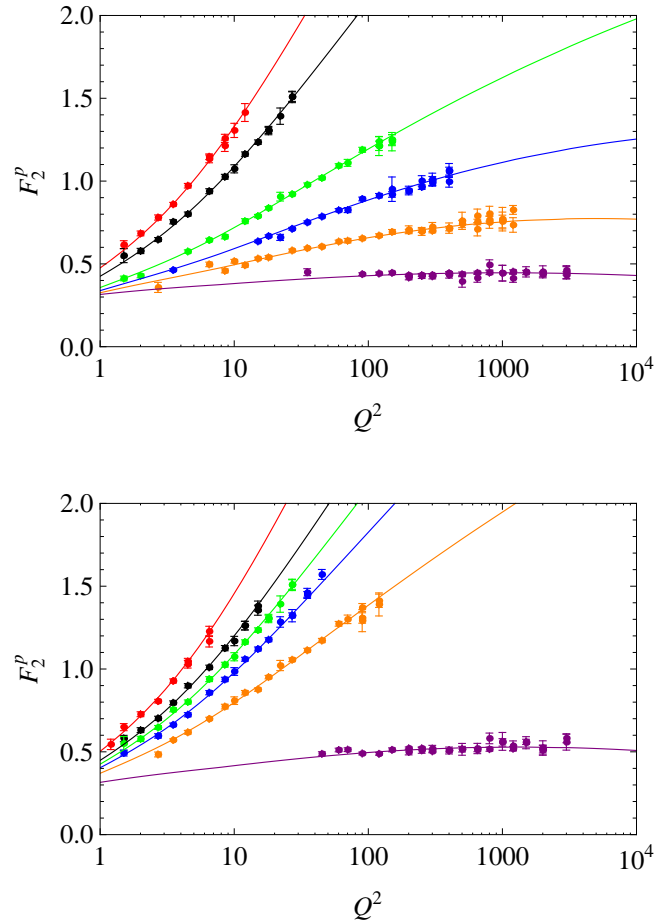 ,width=4.2in }}
\end{center}
\caption{Plots of our fit to $F_2^{\gamma p}$ as a function of $Q^2$ at fixed $x$, compared to the corresponding HERA data \cite{HERAcombined}. Top panel, top to bottom:  $x=0.0002$ (red), 0.0005 (black), 0.0032 (green), 0.008 (blue), 0.02 (orange), 0.08 (purple); Bottom panel, top to bottom: $x=0.00013$ (red), 0.00032 (black), 0.0005 (green), 0.0008 (blue), 0.002 (orange), 0.05 (purple). } \label{figure:F2pQsq}
\end{figure}

\begin{figure} [htb]
\begin{center}
\mbox{\epsfig{file=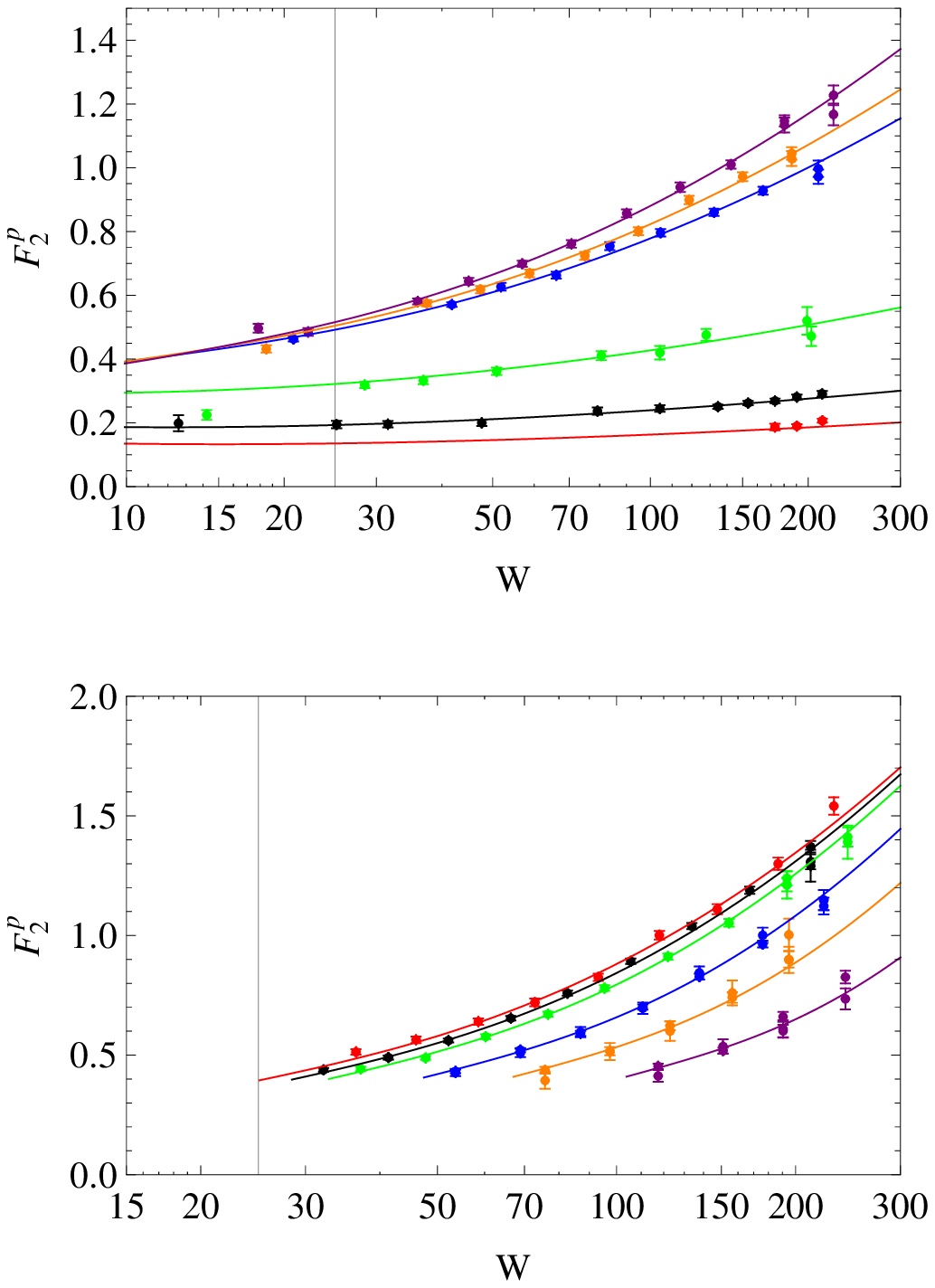,width=4.2in }}
\end{center}
\caption{Plots of the fitted proton structure function
$F_2^{\gamma p}(W,Q^2)$ versus $W$ for representative values of $Q^2$. The vertical lines indicate the cutoff used in the fit, $W\ge25$ GeV.  In the upper panel, top to  bottom: $Q^2=0.15$ (red), $0.25$ (black), $0.65$ (green), $3.5$ (blue), $4.5$ (orange), and $6.5$ (purple) GeV$^2$  In the lower panel:  top to bottom: $Q^2=35$ (red), $90$ (black), $120$ (green), $250$ (blue), $500$ (orange), and $1200$ (purple) GeV$^2$.  The curves in this panel extend in $W$ only to the minimum value allowed by the condition $x\le0.1$.  All data at the specified values of $Q^2$  are shown. } \label{figure:F2pW}
\end{figure}

Plots of the fitted $F_2^{\gamma p}(W,Q^2)$  are shown as functions of $Q^2$ and $W$, along with the HERA data, for representative values of  $x$ and $Q^2$ in Figs. \ref{figure:F2pQsq} and \ref{figure:F2pW}. The fits are clearly very good as is reflected by the excellent overall $\chi^2$ and fit probability. The curves in Fig. \ref{figure:F2pQsq} also extend smoothly and reasonably to values of $Q^2\gtrsim 10^4$ GeV$^2$ above those measured, but needed in our later calculations of neutrino cross sections. 

In the  less conventional plots of $F_2^{\gamma p}$ in Fig. \ref{figure:F2pW}, the effects of the restrictions $W\ge 25$ GeV and $x\le0.1$ are clear. All the data at the given values of $Q^2$ are shown, with the $W$ cutoff indicated in each figure by the vertical line, and the solid curves in Fig. \ref{figure:F2pW} stopping at the lowest $W$ for which $x<0.1$. It is again obvious that the model describes the HERA combined data very well, even, in the upper panel, for $W<25$ GeV.  We note finally that the results are changed very little by increasing the $W$ cutoff to $W=30$ or 35 GeV.

\section{Applications \label{sec:applications}}

\subsection{$\gamma^* p$ cross sections from the fit \label{subset:sigma_gamma*p}}

\begin{figure} [htb]
\begin{center}
\mbox{\epsfig{file=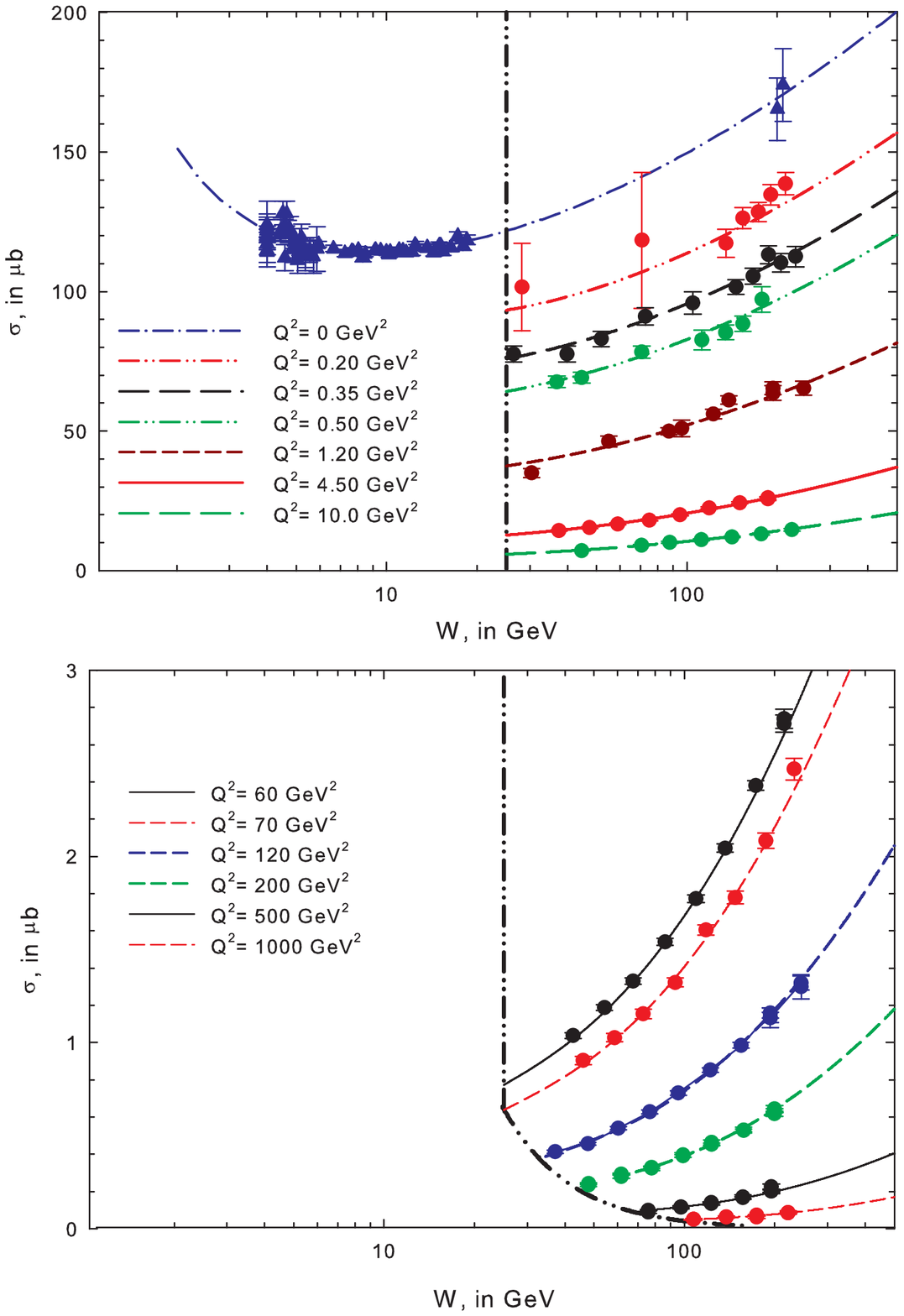,width=4.2in }}
\end{center}
\caption[]{Plots of the $\gamma^*  p$ cross section $\sigma^{\gamma^*p}$, in $\mu$b. vs. $W$, the cms energy of the $\gamma^*p$ system. The upper curve is  for $Q^2$ = 0 to 10  GeV$^2$; the lower curve is for $Q^2=60$ to 1000 GeV$^2$. Notice the very different scales of the vertical axes (cross sections) of the two curves. The  circles which are plotted for Q$^2\ne 0$  are $\gamma^*p$ cross section  data from HERA DIS (deep inelastic scattering) that satisfy the cuts  $W\ge 25$ GeV and $x\le 0.1$; the  $W$ cut for the curves is indicated by the thick dot-dot-dashed vertical line in the upper plot and the thick  dot-dot-dashed boundary  in the lower plot . The plotted cross section  curves of $\sigma^{\gamma^*p}$ for $Q^2>0$ are the sum of  the asymptotic cross section plus the valence cross section. For $Q^2=0$, the curve is the sum of the asymptotic DIS cross section plus the rapidly decreasing Regge-like term used in the Block-Halzen fit to real $\gamma p$ data  \cite{BHgamma-p};  the data for $W>2$ GeV are shown as (blue) triangles.}.  \label{figure:gamP-sigma}
\end{figure}

It is straightforward to evaluate the $\gamma^* p$ cross section, \eq{sigma_F2_relation}, using $F_2^{\gamma p}(W,Q^2)$ from the fit. In Fig.  \ref{figure:gamP-sigma}, we show plots of the $\gamma^* p$ cross sections vs. $W$, the center of mass energy of the $\gamma^*p$ system, for representative values of $Q^2$ up to 10  GeV$^2$ in the upper curves, and  $Q^2=60$ to 1000 GeV$^2$ in the lower curves. The upper curve for $Q^2=0$ is the Block-Halzen fit 2 \cite{BHgamma-p} to the real $\gamma p$ cross section, shown with the all of the data for $W\ge 2$ GeV. Because of the constraints we have imposed  on our fit parameters in \eq{constraints}, using the Block-Halzen parameters as input, our results agree exactly with the asymptotic part of the of the Block-Halzen fit at $Q^2=0$. The added Regge-like term is only 3.5\% of the cross section at $W=25$ GeV, and decreases as $1/W$ for larger $W$, so is essentially negligible on the scale of the figure.

It is evident from this figure, plotted on a logarithmic scale in $W$, that the data follow the quadratic curves in $\ln{W}$ given by our asymptotic form for $F_2^{\gamma p}$, \eq{F2asymp}, with the shapes changing smoothly as a function of  $Q^2$ and approaching the $\gamma p$ cross section for $Q^2\rightarrow 0$. This indicates that the Froissart type behavior characteristic of high-energy hadronic interactions and $\gamma p$ scattering persists experimentally into the region of virtual $\gamma^*p$ scattering, as argued in \cite{bbt2,bhm,bdhmapp} and discussed in more detail in \cite{bdhmFroissart}. A direct calculation of the $\gamma p$ cross section using a parton-level description \cite{HDGPS} to account for the rise in the cross section with increasing $W$ in fact demonstrates the onset of the Froissart type behavior theoretically in that case.


\subsection{$e^{\pm}p$ neutral current cross sections \label{subset:eptotal_cross_sections}}

The integrated $e^{\pm} p$ neutral current (NC) cross sections are of potential interest for $ep$ collider experiments. These involve
integrals over the doubly differential cross section \cite{epCS1, epCS2}
\begin{equation}
\frac{d^2\sigma^{e^{\pm}p}_{NC}}{dxdQ^2}(E_{e},Q^2,x) =
  \frac{2 \pi \alpha^{2} }{Q^{4}}\frac{1}{x}
  [Y_{+} \tilde{F_{2}}(x,Q^{2})
  \mp Y_{-} x\tilde{F_{3}}(x,Q^{2})
  - y^{2}\tilde {F_{L}}(x,Q^{2})],
\label{epCS}
\end{equation}
where $\alpha$ is the fine-structure constant,
$Y_{\pm} = 1 \pm (1 - y)^{2}$, $y=Q^2/2mE_e = Q^2/2e\cdot p$, and 
$\tilde{F_{2}}(x,Q^{2})$, $\tilde{F_{3}}(x,Q^{2})$ and
$\tilde{F_{L}}(x,Q^{2})$
are generalized structure functions which include $Z$ as well as $\gamma$ exchanges between the incident electron and the proton. Next-to-leading (NLO) order QCD calculations predict that the contribution of the longitudinal structure function, $\tilde {F_L}$, to $d^2\sigma /dx dQ^2$ is  $<1\%$  \cite{epCS1}.

The generalized structure functions can be split into terms depending on $\gamma$
exchange ($F_2^{\gamma}$), $Z$ exchange ($F_2^Z$, $xF_3^Z$) and $\gamma/Z$ interference
($F_2^{\gamma Z}$, $xF_3^{\gamma Z}$) as
\begin{equation}
    \tilde{F_2} = F_2^{\gamma} - v_e P_{Z} F_2^{\gamma Z} +
  (v_e^2 + a_e^2) {P_{Z}^{2}} F_2^{Z}  ,
\label{eqn:gen_f2}
\end{equation}
\begin{equation}
   x\tilde{F_3} =  - a_e P_{Z} xF_3^{\gamma Z} + 2 v_e a_e
{P_{Z}^{2}} xF_3^{Z}.
\label{eqn:gen_xf3}
\end{equation}
The standard model predictions for vector and axial vector couplings of the electron to the $Z$ boson are $v_{e} = -1/2 + 2\sin^2\theta_W$ and $a_{e} = -1/2$, where $\theta_W$ is the
Weinberg angle; we use $\sin^2\theta_W=0.231$. The relative fraction of events coming from $Z$ exchange relative to $\gamma$ exchange is given at fixed $Q^2$ by \cite{epCS2}
\begin{equation}
P_{Z}=\frac{1}{\sin^2{2\theta_W}} \frac{Q^{2}}{M_{Z}^{2}+Q^{2}} .
  \label{eqn:Pz}
\end{equation}

The structure functions can be written at the bare quark level, before QCD corrections, in terms of the sum and differences of the quark and anti-quark momentum
distributions,
\begin{equation}
[F_2^{\gamma},F_2^{\gamma Z},F_2^{Z}] =
\sum _q [e_{q}^{2}, 2e_{q}v_{q},v_{q}^{2}+a_{q}^{2}]
 x (q + \bar{q}),
\label{eqn:struc1}
\end{equation}
\begin{equation}
[xF_3^{\gamma Z},xF_3^{Z}] =
\sum _q [e_{q}a_{q},v_{q}a_{q}]
 2x (q - \bar{q}),
\label{eqn:struc2}
\end{equation}
where the sum runs over all quark flavors except the top quark which is too massive to contribute significantly in the region of interest;  $e_{q}$ is the electric charge of the quark and $v_{q}$ and $a_{q}$ are the respective vector and axial couplings of the quark $q$ to the $Z$ boson. For $q=u$, $c$, and $t$, the Standard Model values of the respective vector and axial couplings are $v_{q} = 1/2 -4/3\sin^2\theta_W$ and $a_{q} = 1/2$.  For $q=d$, $s$, and $b$, we have $v_{q} = -1/2 +2/3\sin^2\theta_W$ and $a_{q} = -1/2$.

Following the procedures discussed in \cite{bdhmFroissart}, we can re-express the structure functions given at the quark level in \eq{eqn:struc1} and \eq{eqn:struc2}, in terms of $F_2^{\gamma p}$, a valence term $U$, and a set of non-singlet quark distributions $T_8$, $T_{15}$, and $T_{24}$ which can be determined from $F_2^{\gamma p}$ with minimal input. We have $F_2^{\gamma}=F_2^{\gamma p}$ in Eqs.\ (\ref{eqn:gen_f2}) and (\ref{eqn:gen_xf3}), 
\ba
\label{F2gZ}
F_2^{\gamma Z} &=& (3-\frac{20}{3}\sin^2{\theta_W})F_2^{\gamma p} - \frac{4}{27}\sin^2{\theta_W}T_8-\frac{2}{9}\sin^2{\theta_W}U,\qquad n_f=3, \nonumber \\
F_2^{\gamma Z} &=& (3-\frac{36}{5}\sin^2{\theta_W})F_2^{\gamma p} + \frac{16}{135}\sin^2{\theta_W} \left(T_{15}-T_8\right)-\frac{2}{9}\sin^2{\theta_W}U,\qquad n_f=4, \\
F_2^{\gamma Z} &=& (3-\frac{76}{11}\sin^2{\theta_W})F_2^{\gamma p}  - \frac{8}{297}\sin^2{\theta_W}\left(5T_8-5T_{15}+3T_{24}\right)-\frac{2}{9}\sin^2{\theta_W}U,\qquad n_f=5, \nonumber
\ea
with $F_2^{\gamma p}$ the measured (fitted) DIS structure function, and
\ba
F_2^{Z} &=& (\frac{9}{4}-4\sin^2{\theta_W}+4\sin^4{\theta_W})F_2^{\gamma p} - (9 -8 \sin^2{\theta_W})(\frac{1}{72}T_8+\frac{1}{48}U),\qquad n_f=3, \nonumber \\
F_2^{Z} &=&(\frac{9}{5}-\frac{18}{5}\sin^2{\theta_W}+4\sin^4{\theta_W})F_2^{\gamma p} + (9 -8 \sin^2{\theta_W}) (\frac{1}{90}\left(T_{15}-T_8\right)-\frac{1}{48}U),\qquad n_f=4, \label{F2Z} \\
F_2^{Z} &=& (\frac{45}{22}-\frac{41}{11}\sin^2{\theta_W}+4\sin^4{\theta_W})F_2^{\gamma p} - (9 -8 \sin^2{\theta_W}) (\frac{1}{396}\left(5T_8-5T_{15}+3T_{24}\right)-\frac{1}{48}U),\qquad n_f=5. \nonumber
\ea

The $xF_3$ terms involve only the valence distribution $U$ with
\be
\label{F3terms}
F_3^{\gamma Z}=\frac{3}{2}U \,\, , \,\, F_3^Z=(\frac{3}{4}-\frac{5}{3}\sin^2{\theta_W})U;
\ee
their contributions to the total $e^ \pm p$ neutral current cross sections are negligible.
The structure function $\tilde {F_L}$, which is zero in LO, is given in NLO  by
\be
 \label{FLNLO}
x^{-1}\tilde {F_L}(x,Q^2) = \frac{\alpha_s}{2\pi} C_{Lq}\otimes(x^{-1}F_{20}^{\gamma})+\frac{\alpha_s}{2\pi}2n_f\,C_{Lg}\otimes g. \\
\ee
The details are discussed in \cite{bdhmneutrino}.

The $\gamma/Z$  interference term and the pure $Z$ exchange term will give small contributions to the complete $e^ \pm p$ cross section, so it is sufficient  to estimate them using the ``wee-parton'' approximation in which the quark distributions are taken as equal at small $x$. This is equivalent to setting  the $T_i$ and $U$ terms in Eqs. (\ref{F2gZ})--(\ref{F2Z}) equal to zero \cite{bdhmneutrino}. The $\gamma Z$ interference term and the $Z$ exchange term are then expressed simply in terms of numerical multiples of $F_2^{\gamma p}$, which is known. The longitudinal structure function $\tilde {F_L}$ can be treated similarly.

With this input, it is straightforward to evaluate the NC $e^ \pm p$ cross section by integrating the expression in \eq{epCS} over $x$ and $Q^2$. The $\gamma$ exchange contribution diverges as $1/Q^2$  for $Q^2\rightarrow 0$; we use a lower bound on the $Q^2$ integration $Q_0^2=1$ GeV$^2$. In Table \ref{table:epCS}, we show the resulting values of the NC $e^ \pm p$ cross section over the energy range from $E_e=10^6$ GeV up to $E_e=10^{12}$ GeV, corresponding to center-of-mass energies in the $ep$ system from 1.4 TeV to 1400 TeV.  Contributions from $F_2^{\gamma}$ , $Z$ exchange $F_2^Z$, $\gamma/Z$ interference $F_2^{\gamma Z}$, and $\tilde {F_L}$, to the total $e^ \pm p$ cross section are also given in the Table. Contributions from the $F_3$ terms and the valence term $U$ are very small so can be ignored. As expected, dominant contributions are from the $F_2^{\gamma}$ term.

\begin{table}[ht]                   
%
\begin{center}
     \caption{Neutral current $e^ \pm p$ cross section for $Q^2>1$ GeV$^2$,  in cm$^2$, as a function of $E_e$, the laboratory energy, in GeV.  $\sigma^{e^{\pm}p}$ is the total neutral current cross sections.   $\sigma_{F_2}$, $\sigma_{F_L}$, $\sigma_{F_2^{\gamma Z}}$, and  $\sigma_{F_2^{Z}}$ are the contributions from  $F_2^\gamma$, $\tilde {F_L}$, $F_2^{\gamma Z}$, and $F_2^{Z}$  respectivly.  \label{table:epCS}
 }
   \renewcommand{\arraystretch}{1.5}
   \begin{tabular}{|c|c|c|c|c|c|c|c|}
 \hline

      $E_e$ (GeV)    & $\sigma^{e^{\pm}p}({\rm cm^2})$ & $\sigma_{F_2}({\rm cm^2})$  & $\sigma_{F_L}({\rm cm^2})$  & $\sigma_{F_2^{\gamma Z}}({\rm cm^2})$ & $\sigma_{F_2^Z}({\rm cm^2})$ \\  \hline \hline

           $10^6$    & $2.59\times 10^{-30}$ & $2.12\times 10^{-30}$ & $1.82\times 10^{-32}$ & $5.90\times 10^{-32}$ & $3.88\times 10^{-31}$ \\

            $10^7$    & $3.91 \times 10^{-30}$ & $3.22\times 10^{-30}$ & $2.17\times 10^{-32}$ & $8.80 \times 10^{-32}$ & $5.84 \times 10^{-31}$ \\

            $10^8$    & $5.70\times 10^{-30}$  & $4.73\times 10^{-30}$ & $2.51\times 10^{-32}$ & $1.23 \times 10^{-31}$ &$8.21 \times 10^{-31}$ \\

            $10^9$    & $8.02 \times 10^{-30}$ & $6.73\times 10^{-30}$ & $2.83\times 10^{-32}$ & $1.64 \times 10^{-31}$ & $1.10 \times 10^{-30}$ \\

            $10^{10}$  & $1.10\times 10^{-29}$ & $0.93\times 10^{-29}$ & $3.13\times 10^{-32}$ & $2.11 \times 10^{-31}$ & $1.42 \times 10^{-30}$ \\

            $10^{11}$   & $1.46\times 10^{-29}$ & $1.25\times 10^{-29}$ & $3.43\times 10^{-32}$ & $2.64\times 10^{-31}$ & $1.78\times 10^{-30}$ \\

            $10^{12}$   & $1.91\times 10^{-29}$ & $1.65\times 10^{-29}$ & $3.72\times 10^{-32}$ & $3.23 \times 10^{-31}$ & $2.18\times 10^{-30}$ \\

%
%
%
%

            \hline

        \end{tabular}
        \renewcommand{\arraystretch}{1}
        \end{center}
\end{table}
%

\begin{figure} [ht]
\begin{center}
\mbox{\epsfig{file=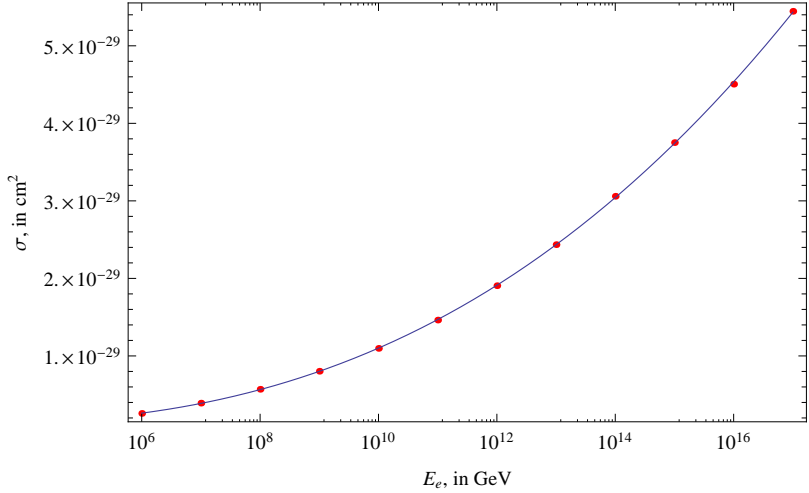,width=4.2in }}
\end{center}
 \caption{Plots of $e^ \pm p$ NC cross section for $Q^2>1$ GeV$^2$, in cm$^2$, vs. $E_e$, the laboratory neutrino energy, in GeV.  The points are the numerical calculations of Table \ref{table:epCS}.
 } \label{fig:epCS}
\end{figure}

The integration of the doubly differential cross section in \eq{epCS} over $x$ and $Q^2$ needed to obtain these results involves a factor $dx/x=d(\ln{x})$ from the pre factor $1/x$ in \eq{epCS}, with an upper limit on $x$ of $2mE_e/Q_{\rm min}^2$. The $Q^2$ integration converges rapidly.  As a result,  a Froissart bounded $F_2^{\gamma p}$ which behaves asymptotically as $\ln^2(1/x)$ for decreasing $x$ leads to  an integrated $e p$ cross section that grows asymptotically as $\ln^3E_e$. This modified Froissart behavior for the integrated cross sections was originally noted  in the case of neutrino-proton scattering in \cite{kniehl}, where the bound on the integrated cross section is proportional to $\ln^3E_\nu$, and was re-emphasized for that case in  \cite{bdhmneutrino}.\footnote{The authors of \cite{kniehl} state incorrectly that the work in \cite{bbmt,bhm} claims that the $\nu N$ cross section rises only as $\ln^2E_\nu$ in the limit of large $E_\nu$. Those references actually assume the Froissart saturated form $\ln^2(1/x)$ only for the structure function $F_2^{\gamma p}$ and the corresponding neutrino structure function $F_2^{\nu N}$, and not for the integrated lowest-order weak $\nu N$ cross section. This confusion between the virtual gauge boson-$N$ and total $\nu N$ cross sections was clarified in \cite{bhm11}.} More generally, the integrated $e^{\pm}p$  and $\nu N$ cross sections should behave asymptotically as $\ln^3E_e$  or $\ln^3E_\nu$,  with sub-dominant terms which behave as lower powers of $\ln{E_e}$ or $\ln{ E_\nu}$. We have therefore done a 4 parameter fit to the $e^{\pm}p$ cross section of the form  $\sigma^{e^{\pm}p}= a+b\ln E_e+c \ln E_e^2+d\ln E_e^3$, using the data in Table \ref{table:epCS} and its extension to higher $E_e$, to obtain the analytic expression for the integrated cross section for $Q^2>1$ GeV$^2$,
\be
\sigma^{e^{\pm}p}_{NC}(E_e)=3.058 \times 10^{-30}-3.593  \times 10^{-31} \ln E_e
+1.339  \times 10^{-32} \ln^2 E_e+7.472  \times 10^{-34} \ln^3 E_e. \label{epCSanalytic}
\ee
Here $E_e$ is in GeV and the constants and cross section in cm$^2$. We plot the data from Table \ref{table:epCS} and the fitted cross section in Fig. \ref{fig:epCS}. The $\ln^3 E_e$  parameterization is excellent, with numerical agreement better than 1 part in 1000. Detection of this behavior in $e^{\pm}p$ scattering would be a clear indication of a Froissart boundedness of the underlying $\gamma^*p$ cross section.

\subsection{Neutrino-nucleon cross sections}

The charged current (CC) and neutral current (NC) cross sections for the scattering of neutrinos and antineutrinos on an isoscalar nucleon target $N=(p+n)/2$ at ultra-high energies have been calculated by a number of authors using different approaches. The results depend on the behavior of structure functions at ultra-small $x$, down to $x\sim 10^{-12}$ for the neutrino energies $E_\nu\sim 10^{16}$ GeV that will potentially be accessible at  cosmic ray neutrino detectors now operating  (ICECUBE \cite{icecube}, Baikal \cite{baikal}, ANTARES \cite{antares}, HiRes \cite{HiRes}, AUGER \cite{auger}), or under development  (ARA \cite{ara}, ARIANNA \cite{ARI}) or proposed (JEM-EUSO \cite{euso,euso2}).  

The cross section calculations of \cite{reno-quigg,fmr1,gandhi96,gandhi98,c-ss,gluck,c-sms}, which use parton distributions derived in analyses of DIS based on the DGLAP evolution equations \cite{dglap1,dglap2,dglap3}, require the extrapolation of those power-law  dominated parton distributions far outside the experimental region, with results that become increasingly uncertain at very high neutrino energies. Alternative approaches such as that of Fiore {\em et al.} \cite{fiore1,fiore2}  emphasize specific ideas about the behavior of structure functions at small $x$ which go beyond the usual DGLAP approach, and allow for parton recombination effects and the damping of structure functions at very small $x$ \cite{GLRsmxqcd,desy1990}. 

In the approach adopted here, we emphasize instead the expression of the neutrino cross sections directly in terms of a Froissart-bounded $F_2^{\gamma p}$ following \cite{ bbmt,bhm}, with the inclusion of small corrections not considered there. We do not introduce a specific mechanism for the boundedness within the general Froissart framework. The calculations are discussed  in detail in \cite{bdhmFroissart} and \cite{bdhmneutrino}. We will not repeat the details or the arguments for this approach here, but simply present the results of updated calculations in which the fit to the HERA data on $F_2^{\gamma p}$ given in \cite{bhm,bdhmapp} is replaced by the fit constructed here.  

\subsubsection{Numerical evaluation of the CC and NC $\nu N$ cross sections}

\begin{figure} [ht]
\begin{center}
\mbox{\epsfig{file=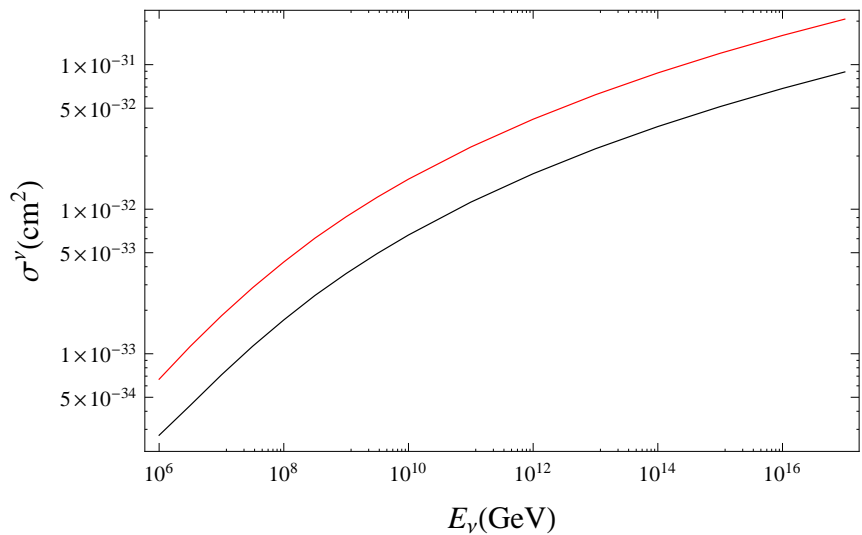,width=4.5in }}
\end{center}
\caption[]{Plots of the $\nu N$ cross sections, in cm$^2$, versus  $E_\nu$, the laboratory neutrino energy, in GeV,  calculated using the  extrapolation of the global fit to the HERA data on $F_2^{\gamma p}(x,Q^2)$ to small $x$,  and the relations between $F_2^{\nu(\bar{\nu})}$, $F0_2^{\nu(\bar{\nu})}$, and $F_2^{\gamma p}$  with  NLO treatments of the small functions $T'_i$ and of the subdominant structure functions $F_3^{\nu(\bar{\nu})}$ and $F_L^{\nu(\bar{\nu})}$ discussed in \cite{bdhmneutrino}.  The upper curve (red) is the CC cross section and lower curve (black) is the NC cross section.}
\label{fig:heraCC-NC}
\end{figure}

In Fig. \ref{fig:heraCC-NC} we show the $\nu N$   CC and NC cross sections, in cm$^2$, as a function of $E_\nu$, in GeV, for large $E_\nu$, calculated using our fit to  $F_2^{\gamma p}$. 

\begin{table}[ht]                   
%
\begin{center}
     \caption{Charged current $\nu N$ cross sections,  in cm$^2$, as a function of $E_\nu$, the laboratory neutrino energy, in GeV. Here  $\sigma_{CC}$ is the total charged current cross section.   $\sigma_{F_2}$ , $\sigma_{F_3}$,  $\sigma_{F_L}$, and $\sigma_U$ are the contributions from $F_2^\nu$,  $F_3^\nu$, $F_L^\nu$ , and from the valence quark distribution $U$.  \label{table:CCs}
 }
   \renewcommand{\arraystretch}{1.5}
   \begin{tabular}{|c|c|c|c|c|c|c|c|}
 \hline
      $E_\nu$ (GeV)    & $\sigma_{CC}({\rm cm^2})$ & $\sigma_{F_2}({\rm cm^2})$  & $\sigma_{F_3}({\rm cm^2})$  & $\sigma_{F_L}({\rm cm^2})$ & $\sigma_{U}({\rm cm^2})$ \\  \hline \hline

           $10^6$    & $6.66\times 10^{-34}$ & $6.97\times 10^{-34}$ & $0.01\times 10^{-34}$ & $-0.15\times 10^{-34}$ & $-0.16\times 10^{-34}$ \\

            $10^7$    & $1.83 \times 10^{-33}$ & $1.89\times 10^{-33}$ & $0.00$ & $-0.03 \times 10^{-33}$ & $-0.03 \times 10^{-33}$ \\

            $10^8$    & $4.31\times 10^{-33}$  & $4.42\times 10^{-33}$ & $0.00$ & $-0.07 \times 10^{-33}$ &$-0.04 \times 10^{-33}$ \\

            $10^9$    & $8.87 \times 10^{-33}$ & $9.00\times 10^{-33}$ & $0.00$ & $-0.14 \times 10^{-33}$ & $0.00$ \\

            $10^{10}$  & $1.61\times 10^{-32}$ & $1.64\times 10^{-32}$ & $0.00$ & $-0.02 \times 10^{-32}$ & $0.00$ \\

            $10^{11}$   & $2.69\times 10^{-32}$ & $2.73\times 10^{-32}$ & $0.00$ & $-0.04\times 10^{-32}$ & $0.00$ \\

            $10^{12}$   & $4.19\times 10^{-32}$ & $4.24\times 10^{-32}$ & $0.00$ & $-0.05 \times 10^{-32}$ & $0.00$ \\

            $10^{13}$   & $6.19\times 10^{-32}$ & $6.26\times 10^{-32}$ & $0.00$ & $-0.07\times 10^{-32}$ & $0.00$ \\

            $10^{14}$   & $8.77\times 10^{-32}$ & $8.85\times 10^{-32}$ & $0.00$ & $-0.09\times 10^{-32}$ & $0.00$ \\

            $10^{15}$   & $12.0 \times 10^{-32}$ & $12.1 \times 10^{-32}$ & $0.00$ & $-0.11 \times 10^{-32}$ & $0.00$ \\

            $10^{16}$   & $15.9 \times 10^{-32}$ & $16.1\times 10^{-32}$ & $0.00$ & $-0.13\times 10^{-32}$ & $0.00$ \\

            $10^{17}$   & $20.7 \times 10^{-32}$ & $20.8\times 10^{-32}$ & $0.00$ & $-0.16\times 10^{-32}$ & $0.00$ \\

            \hline

        \end{tabular}
        \renewcommand{\arraystretch}{1}
        \end{center}
\end{table}
%

\begin{table}[ht]                   
%
\begin{center}
     \caption{Neutral current $\nu N$ cross sections,  in cm$^2$, as a function of $E_\nu$, the laboratory neutrino energy, in GeV:  $\sigma_{NC}$ is the total neutral current cross section.    $\sigma_{F_2}$ , $\sigma_{F_3}$,  $\sigma_{F_L}$, and $\sigma_U$ are the contributions from $F_2^\nu$,  $F_3^\nu$, $F_L^\nu$ , and from the valence quark distribution $U$. \label{table:NCs}
 }
   \renewcommand{\arraystretch}{1.5}
   \begin{tabular}{|c|c|c|c|c|c|c|c|}
 \hline

      $E_\nu$ (GeV)    & $\sigma_{NC}({\rm cm^2})$ & $\sigma_{F_2}({\rm cm^2})$  & $\sigma_{F_3}({\rm cm^2})$  & $\sigma_{F_L}({\rm cm^2})$ & $\sigma_{U}({\rm cm^2})$ \\  \hline \hline

           $10^6$    & $2.73\times 10^{-34}$ & $2.75\times 10^{-34}$ & $0.00$ & $0.05\times 10^{-34}$ & $-0.07\times 10^{-34}$ \\

            $10^7$    & $7.14 \times 10^{-34}$ & $7.59\times 10^{-34}$ & $0.00$ & $-0.31 \times 10^{-34}$ & $-0.13 \times 10^{-34}$ \\

            $10^8$    & $1.71\times 10^{-33}$  & $1.81\times 10^{-33}$ & $0.00$ & $-0.08 \times 10^{-33}$ &$-0.02 \times 10^{-33}$ \\

            $10^9$    & $3.59 \times 10^{-33}$ & $3.74\times 10^{-33}$ & $0.00$ & $-0.16 \times 10^{-33}$ & $0.00$ \\

            $10^{10}$  & $6.63\times 10^{-33}$ & $6.90\times 10^{-33}$ & $0.00$ & $-0.26 \times 10^{-33}$ & $0.00$ \\

            $10^{11}$   & $1.12\times 10^{-32}$ & $1.16\times 10^{-32}$ & $0.00$ & $-0.04\times 10^{-32}$ & $0.00$ \\

            $10^{12}$   & $1.76\times 10^{-32}$ & $1.82\times 10^{-32}$ & $0.00$ & $-0.06 \times 10^{-32}$ & $0.00$ \\

            $10^{13}$   & $2.62\times 10^{-32}$ & $2.70\times 10^{-32}$ & $0.00$ & $-0.08\times 10^{-32}$ & $0.00$ \\

            $10^{14}$   & $3.73\times 10^{-32}$ & $3.83\times 10^{-32}$ & $0.00$ & $-0.10\times 10^{-32}$ & $0.00$ \\

            $10^{15}$   & $5.12 \times 10^{-32}$ & $5.26 \times 10^{-32}$ & $0.00$ & $-0.13 \times 10^{-32}$ & $0.00$ \\

            $10^{16}$   & $6.84 \times 10^{-32}$ & $7.00\times 10^{-32}$ & $0.00$ & $-0.16\times 10^{-32}$ & $0.00$ \\

            $10^{17}$   & $8.91 \times 10^{-32}$ & $9.10\times 10^{-32}$ & $0.00$ & $-0.19\times 10^{-32}$ & $0.00$ \\

            \hline

        \end{tabular}
        \renewcommand{\arraystretch}{1}
        \end{center}
\end{table}
%

 The values of the CC and NC cross sections  over the energy range from $10^6$ GeV up to $10^{17}$ GeV are shown in Table \ref{table:CCs} and Table \ref{table:NCs} where contributions from $F_2^{\nu}$ , $F_3^{\nu(\bar{\nu})}$, $F_L^{\nu(\bar{\nu})}$, and the valence quark distribution $U$, to the total cross sections are also shown. Note that the contribution from  $F_3^{\nu(\bar{\nu})}$ to the charged current neutrino cross section is very small and decreases as $E_\nu$ increases.


\subsubsection  {Implications of high energy weak $\nu N$ scattering  for strong hadronic interactions}

\begin{figure} [ht]
\begin{center}
\mbox{\epsfig{file=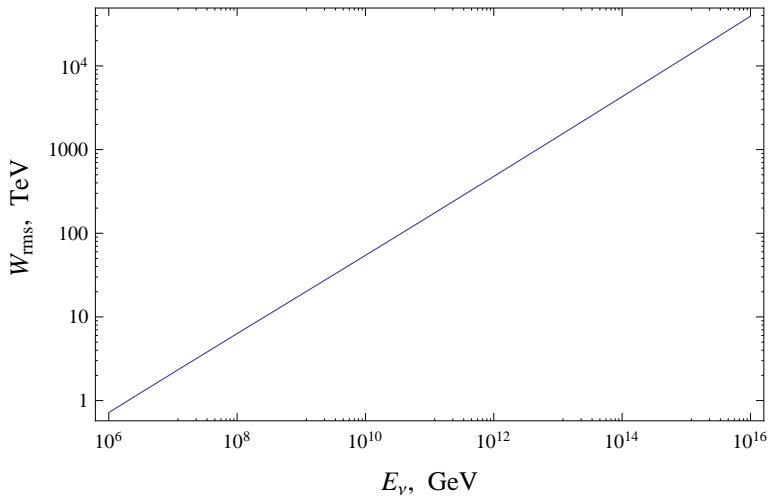
,width=4.2in }}
\end{center}
 \caption{Plot of $W_{{\rm rms}}$, the root mean square hadronic cms energy of the virtual vector boson $-N$ system,  in TeV, versus $E_\nu$, the laboratory neutrino energy for CC scattering, in GeV.  
 } \label{fig:Wrms}
\end{figure}

Measurement of these new high energy neutrino cross section  have the potential of allowing measurements of  {\em hadronic interactions} and tests of the underlying theory at heretofore undreamt of high energies. In Fig. \ref{fig:Wrms}, we plot $W_{\rm rms}$,    the CC root mean square cms energy of the strong interaction of the virtual vector boson, in TeV,  against the energy $E_\nu$ of the laboratory neutrino, in GeV. We see in Fig \ref{fig:Wrms} that for $E_\nu\approx 10^{10}$ GeV that  we have already reached the strong interaction energy of $ 57$ TeV, the  cms energy in  $pp$ collisions reached by the Pierre Auger Collaboration \cite{POAp-air} using cosmic ray protons with $E_p\sim 1.7\times 10^{18}$ eV, near the expected Greisen-Zatsepi-Kuzmin  limit on the incident proton spectrum \cite{greisen,ZKlimit}. The incident neutrino spectrum is not subject to this limit, and may well extend to much higher energies. Thus, for $E_\nu\approx 10^{10},\,10^{11},\,10^{12},\,10^{13},\,10^{14}$ GeV ($10^{19},\,10^{20},\,10^{21},\,10^{22},\,10^{23}$ eV), we are at a strong interaction cms energies  $W_{\rm rms}\approx 54,\,160,\,480,\,1430$ and 4300 TeV, showing that the detection of  ultra-high energy neutrinos would allow us to reach exceedingly high hadronic energies.

\begin{figure} [ht]
\begin{center}
\mbox{\epsfig{file=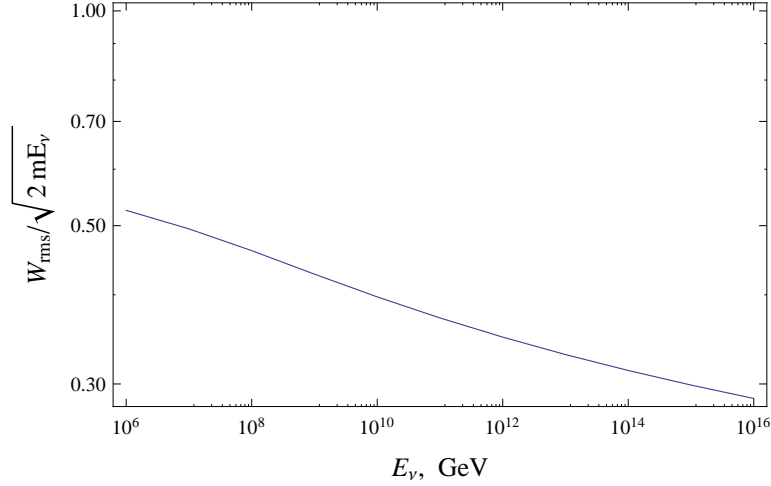,width=4.2in }}
\end{center}
 \caption{Plot of the fraction $W_{{\rm rms}}/\sqrt{2mE_\nu}$ of the cms energy of the $\nu N$ system  which appears in the hadronic cms energy of the virtual vector boson-$ N$ system,   versus $E_\nu$, the laboratory neutrino energy in CC scattering, in GeV.  
 } \label{fig:frac}
\end{figure}

The conversion of the weak $\nu N$ cms energy into the cms energy of the virtual boson$\,N$ system is very  efficient.  In Fig. \ref{fig:frac}, we plot  the fraction of the $\nu N$ cms energy that is in the final hadronic system, defined here as the ratio
\ba
{\rm frac}&=& \left( \frac{\left<s\right>}{2mE_\nu}\right)^{1/2}=\frac{W_{\rm rms}}{\left(2mE_\nu\right)^{1/2}}, \label{cmsWfrac}
\ea
 versus the laboratory neutrino energy $E_\nu$, in GeV. Here $W_{\rm rms}$ is the square root of  $\left< s\right>$, the mean of the square of the cms  hadronic energy of the virtual vector boson - isobaric nucleon $N$ system, and $2mE_\nu$ is the square of the cms energy of the $\nu N$ system. From Fig. \ref{fig:frac}, it is clear that a very large percentage, of the order of 50\%, of the cms energy of the $\nu N$ system is found as hadronic energy for the lower energy neutrinos, decreasing to about 30 \% at the highest energies.

Just as in the $ep$ system, a test for the Froissart bounded behavior of the structure function $F_2^{\gamma p}$ is  in a $\ln^3E_\nu$ asymptotic growth of the  $\nu N$ total cross sections. The present calculations give
\ba
\sigma_{CC}&=&-2.097\times 10^{-32}+4.703\times 10^{-33}\ln{E_\nu}-3.666\times 10^{-34} \ln^2{E_\nu} +1.010\times 10^{-35}\ln^3{E_\nu},\label{CC}\\
\sigma_{NC}&=&\ \ \, 1.021\times 10^{-32} +2.239\times 10^{-33}\ln{E_\nu}-1.700\times 10^{-34}\ln^2{E_\nu}+4.534\times 10^{-36}\ln^3{E_\nu}, \label{NC}
\ea
where the cross sections are in cm$^2$ and $E_\nu$ is in GeV.  We estimate about a 1 to 4 \% error in the above cross sections.

\begin{figure} [ht]
\begin{center}
\mbox{\epsfig{file=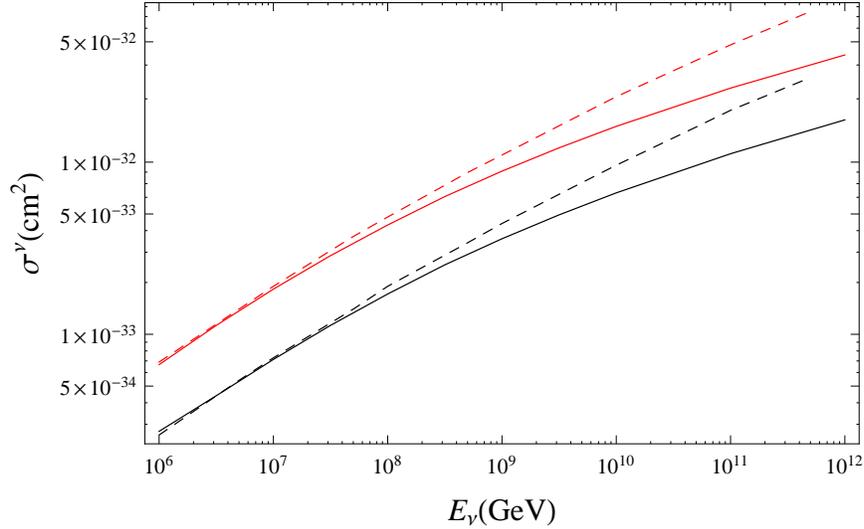,width=4.5in }}
\end{center}
\caption[]{Plots of $\nu N$ cross sections, in cm$^2$, vs. $E_\nu$, the laboratory neutrino energy, in GeV. Our CC cross section is the upper solid (red) curve and our NC cross section is the lower solid (black) curve;  the  CC cross section of CSMS is the upper dashed (red) curve and their NC cross section is the lower dashed (black) curve.  All cross section calculations include the b-quark.}
   \label{fig:compareplot}
\end{figure}

In Fig.\ \ref{fig:compareplot}  we compare our UHE (ultra-high-energy) cross sections from Fig.\ \ref{fig:heraCC-NC} with those of Cooper-Sarkar, Mertsch, and Sarkar (CSMS),  who used the HERA-based PDF set HERAPDF1.5, and  included the $b$ quark but not the $t$ in their computations.  Their quoted error estimates are in the 2\%-4\% range, comparable to ours, when  they exclude those PDF sets which lead to an unacceptably steep rise in the cross section or allow negative values of the gluon PDF at small $x$ and small $Q^2$. Note that our UHE cross sections are very close to the CSMS cross sections in the low energy region ($E_\nu < 10^7$ GeV), eliminating a discrepancy encountered with our previous model for $F_2^{\gamma p}$ \cite{bdhmneutrino}.

The strong divergence of the CSMS and present cross sections at UHE provides a clear distinction between cross sections based on standard parton distributions extrapolated to ultra-small $x$ and the cross sections based on the Froissart bound for strong hadronic processes presented here. They are an additional test of the strong interactions at exceedingly high energies. 

\section{Summary and conclusions \label{sec:conclusions}}
Using a new Froissart-bounded parametrization of the DIS structure function $F_2^{\gamma p}$, we have fitted the the experimental HERA  results on DIS \cite{ZEUS1,ZEUS2,ZEUS3,H1}  in the region $x\le 0.1$ and $0.1\le Q^2\le 50 00$ GeV$^2$, restricted to virtual $\gamma^* p$ center-of- mass energies $W\geq 25$ GeV. We have used the results to calculate the cross section $\sigma^{\gamma*p}(W, Q^2)$ for virtual $\gamma^* p$ scattering, and connected it to the known Froissart-bounded high energy cross section for real $\gamma p$ scattering. 

The new parametrization for $F_2^{\gamma p}$ is divided into two parts, an asymptotic (high energy) part corresponding to $W\ge 25$ GeV and $x\le 0.1$ and a low energy part corresponding to the valence-quark contributions. Our parameterization of $F_2^{\gamma p}$ requires that $F_2^{\gamma p}\rightarrow 0$ as $Q^2\rightarrow 0$. In the fitting procedure, we also require that when $Q^2\rightarrow 0$, the asymptotic portion of $\sigma^{\gamma^*p}(W, Q^2)\rightarrow 0$ go smoothly into the measured asymptotic real $\gamma p$ cross section $\sigma^{\gamma p} (W)$, found by Block-Halzen \cite{BHgamma-p} to be of the form $c_0^{BH} +c_1^{BH}\ln (\nu/m) +c_2^{BH}\ln^2 (\nu/m)$, where $\nu$ is the $\gamma$ laboratory energy. This fixes 3 of the 12 parameters used in the asymptotic part of $F_2^{\gamma p}$. The valence contributions are then  added to obtain the full parametrization of $F_2^{\gamma p}$. A $\chi^2$ fit  to  the 395 HERA datum points with $W\geq 25$ GeV and $x<0.1$ (with 9 free parameters) gives a $\chi^2/{\rm d.o.f.}=0.95$, with a goodness-of-fit probability of 0.76.  

The fit also determines the mass parameter $M$ of \eq{Dfinal}. This mass specifies the most important mass region in the dispersion relation for $F_2^{\gamma p}$ in $Q^2$,  and can be interpreted as the effective mass of the virtual vector boson (or group of bosons) that  interacts strongly with the nucleon. The result is $M= 0.87$ GeV, in the mass region covered by the vector mesons $\rho,\ \omega$ and $\phi$, a result compatible with the idea of vector meson dominance. 

Since the asymptotic portion of $\sigma^{\gamma^*p}(W,Q^2)$ was constructed to be compatible with a Froissart bound of $\ln^2  W$, so is the high energy portion of the fit to $F_2^{\gamma p}(x, Q^2)$.  In particular,  for small $x$ and all $Q^2\geq 0$, $F_2^{\gamma p}(x, Q^2)$ is bounded by $\ln^2(1/x)$.

We use our new results on $F_2^{\gamma p}$ to calculate the $ep$ cross section, and update earlier calculations \cite{bdhmneutrino} of $\nu N$ cross sections. For high  initial energies of either the $ep$ or $\nu N$ collisions, the integration over the Froissart-bounded structure function proportional to $\ln^2(1/x)$ needed to obtain the total cross section gives a high energy total cross section $\sigma(E)$ bounded by $\ln^3E$ where $E$ is the laboratory energy of the electron or neutrino, i.e., $\sigma(E)=\alpha +\beta\ln E+\gamma \ln^2 E+\delta\ln^3 E$, providing a new test for the boundedness. In a certain sense, this is a new type of unification of the electromagnetic and weak interactions, brought about by the high energy cross sections for both  being controlled by the Froissart bound on hadronic processes.   In essence, the strong interactions  determine both the weak and electromagnetic cross sections up to factors of the electroweak gauge boson - nucleon couplings.     It is possible that these considerations can be extended to quantum gravitational interactions through the exchange of a spin 2 graviton. The virtual graviton would effectively interact strongly---up to the gravitational coupling---with hadrons. A possible example would be the gravitational interaction of high energy sterile neutrinos  with nucleons.   

We emphasize that measuring  ultra-high energy neutrino $-N$ cross sections would allow us to investigate strong interactions, i.e., the hadronic Froissart bound, at incredibly high energies, opening up new techniques for studying high energy  hadron physics.  For example, if one were able to measure the total cross section for neutrino interactions at $E_\nu=10^{14}$ GeV, it would allow a measurement of the Froissart bound at $\approx 4000$ TeV.


\begin{acknowledgments}

 M.\ M.\ B.\ and L.\ D.\ would like to thank the Aspen Center for Physics, where this work was  supported in part by NSF Grant No. 1066293, for its hospitality. 
 M.\ M.\ B.\  would like to thank Prof.\ Arkady \ Vainshtein  for valuable discussions.   P.\ H.\ would like to thank Towson University Fisher College of Science and Mathematics for support. We would also like to thank Prof. Douglas McKay for his encouragement and early participation.  

\end{acknowledgments}

\bibliography{small_x_references}

\end{document}